# Stochastic Aspect of the Tomographic Reconstruction Problems in a Transport Model

Igor Kharin

*Abstract*. **The stochastic differential and integral equations describing the system of particles weakly interacting among themselves which are absorbed and scattered by particles of a medium are considered. The time-dependent transport equation with scattering is studied taking into account stochastic nature of parameters in nuclear imaging. Using dynamic attenuated Radon transform the solution of transport equation may be derived taking into account of the scattering as perturbation. We analyze the influence of the random variables upon the image reconstruction both generally and in more details for the case of point source.**

**It is shown by the example of the method of the filtered back projection (FBP) that unaccounted small fluctuations of attenuation coefficient can cause essential distortions of image texture and degradation of the resolution at image reconstruction in single-photon emission computerized tomography (SPECT) and less in X-ray computerized tomography (CT). The mechanism of these distortions is analyzed. The way for their elimination is shown for point sources. We demonstrate that for the practical purposes it is enough to define averaged attenuation coefficient in investigated area when its difference from true attenuation coefficient has certain stochastic properties. It is shown that for positron emission tomography (PET) stochastic components in parameters of a transport model without the scattering can be taken into account of the corrections of projections.**

*Keywords*: **Image Reconstruction, Nuclear Imaging, PET, SPECT, Transport Equation, X-ray CT, Radon Transform.**

## I. Introduction

ONE of major problems of image reconstruction from the projection data in a tomography is the problem of noise elimination. The physical sources of a noise can conditionally be sectioned into two basic types. Moreover there are the noise which is bounded with the uncertainty of detection methods, internal noise of electronics etc. This class of the noise sources is "the noise of devices". The second kind of a noise is generated physical processes having stochastic nature such as processes of radiation, absorption, scattering and detection in abstract sense.

In practice the most widespread reconstruction methods are based on analytical methods of the solution of problems in CT. Analytic methods are computationally inexpensive but they can be no effective in the presence of a noise [1]-[3]. In systems of reconstructive tomography the noise in each reconstructed element depends on the noise of each projection passing through the given information element and computational noise which bounds with algorithm of the reconstruction. Thus with usage of integral transformations the change of the noise from the element to element influences upon the reconstruction which depends on convolution core [4] because of this central areas are more corrupted by the noise than peripherals. Statistical techniques of the reconstruction have some attractive features for accounting a noise [3], [5], [6] though sometimes they require prohibitively (for practical applications) long computation times. Besides in some cases the problem becomes more complicated when it is necessary in principle to take into account dependence of parameters on time in dynamic CT [5], on energy as for example in polyenergetic X-ray CT [6]. It is obvious that any reconstruction methods are applied in context of certain physicomathematical model of processes in CT. The consecutive account of different stochastic aspects within the framework of corresponding model is necessary for the creation of effective algorithms. If scatter is included the transport model is most appropriate.

For X-Ray CT the theoretical results with account of scattering [7] and numerical methods for special choices of scattering term [8] were obtained for the models based on transport formulation. In most practical applications of emission computed tomography (ECT) in comparison with X-ray CT there is additional complexity associated with the unknown attenuation coefficient. On the basis of transport equation, it is possible to find the distribution of radiating sources if one formulates the hypothesis about attenuation coefficient $\mu$ : $\mu$ is a constant in known area [9], $\mu$ is a function which is in accord with defined conditions as in [10] and [11]. But useful in practice the definition way of $\mu$ from the emission data has been not found.

The contribution of dispersed particles at the detection of projection data can be reached with the using collimation on energy and angle [12]. At the same time we can do nothing similar with other the noise of second type as some uncertainty of attenuation coefficient is. How in practice it is

I. G. Kharin is with the National Science Center "Kharkov Institute of Physics and Technology", Akademicheskaia 1, Kharkov, 61108 Ukraine, (e-mail: igor@isc.kharkov.com, http://isqom.org/usr/igor)

met besides that we only assume the defined form of $\mu$ dependence on coordinate, energy and time else researching objects are random inhomogeneous mediums as a rule. Moreover the quantum nature of processes is the reason because of which we can not say from what the fluctuation of projection data is depended on at the given concrete measurement: either the density fluctuation of radiation sources or number of emitted photons or attenuation coefficient or scattering parameters etc. Apparently, it will be useful to research whenever possible all aspects influence upon image reconstruction of the fluctuation of parameters for the development of practically useful methods of the solution so the composite problems (both physical and mathematical sense).

In the given paper the reconstruction problems bounded with account of the second type of a noise in SPECT, PET and X-ray CT are studied within the framework of transport model. Here we investigate stochastic aspects of transport model on base of the fundamental kinetic equation for distribution function of particles when we can neglect their interaction with each other. Then transport equation and corresponding integral transformation of the distribution of emitted particles are studied for photons in case of the contribution of fluctuations of parameters. We analyze the influence of the random components upon image reconstruction in CT both generally and in more details for the case of a point source when attenuation coefficient is assumed to have a prior probability distribution, called a Gaussian Markov random field.

## II. Stochastic Aspect of a Transport Model

Let's consider the equation bounding distribution function of particles $I_{\mathbf{p}}(\mathbf{x},t)$ with impulse $\mathbf{p}$, distribution function $F(\mathbf{x},t;\mathbf{p})$ of particles with the impulse $\mathbf{p}$ radiated from source at the point $\mathbf{x}$ in area $\Omega$ and time $t$. Inside $\Omega$ absorption cross-section $\sigma(\mathbf{x},t;E)$ and scattering kernel $W(\mathbf{x},t;\mathbf{p},\mathbf{p}')$ are not equal to zero. In this case it is well known from statistical physics the general kinetic equation is written in the following form

$$\left(\frac{\partial}{\partial t} + \mathbf{v}\frac{\partial}{\partial \mathbf{x}} + \frac{\partial V}{\partial \mathbf{x}}\frac{\partial}{\partial \mathbf{p}}\right) I_{\mathbf{p}}(\mathbf{x},t)$$
$$= F(\mathbf{x},t;\mathbf{p}) - \sigma(\mathbf{x},t;\mathbf{p}) I_{\mathbf{p}}(\mathbf{x},t) + \qquad (1)$$
$$\int \left[ W(\mathbf{x},t;\mathbf{p},\mathbf{p}') I_{\mathbf{p}'}(\mathbf{x},t) - W(\mathbf{x},t;\mathbf{p}',\mathbf{p}) I_{\mathbf{p}}(\mathbf{x},t) \right] d^3 p',$$

where $V$ is interaction of particles each other and external field, $\mathbf{v} = d\mathbf{x}/dt$. Further we will study the systems consisting of medium and particles when the energy interaction $V$ can be neglected, as it is for example in the case of photons or neutrino. Kernel $W(\mathbf{x},t;\mathbf{p},\mathbf{p}')$ and absorption cross-section $\sigma(\mathbf{x},t;\mathbf{p})$ can be parameterized depending on $i^{th}$ component

of the medium

$$W(\mathbf{x},t;\mathbf{p},\mathbf{p}') = \vec{N}(\mathbf{x},t)\vec{w}(\mathbf{p},\mathbf{p}') = \sum_i N_i(\mathbf{x},t)\,\beta_i(\mathbf{p},\mathbf{p}'),$$
$$\sigma(\mathbf{x},t;\mathbf{p}) = \sum_i N_i(\mathbf{x},t)\sigma_i(\mathbf{p}), \qquad (2)$$

where $N_i$, $\sigma_i$ and $\beta_i(\mathbf{p},\mathbf{p}')$ is the density, cross section of the absorption and scattering function for $i^{th}$ component of the medium in $\Omega$ accordingly. Introducing attenuation coefficient $\mu(\mathbf{x},t;\mathbf{p}) = \sigma(\mathbf{x},t;\mathbf{p}) + \int W(\mathbf{x},t;\mathbf{p}',\mathbf{p})\,d^3 p'$ and taking into account the above mentioned we can rewrite (1) in form of the transport equation

$$\left[\frac{\partial}{\partial t} + \mathbf{v}\nabla + \mu(\mathbf{x},t;\mathbf{p})\right] I_{\mathbf{p}}(\mathbf{x},t)$$
$$= F(\mathbf{x},t;\mathbf{p}) + \int W(\mathbf{x},t;\mathbf{p},\mathbf{p}')I_{\mathbf{p}'}(\mathbf{x},t)\,d^3 p'. \qquad (3)$$

Let us study the case when last term in (3) is a small value. And let us suggest that we know how to solve (3) without the scattering term. In this case we will denote this solution as $I_{\mathbf{p}}[0]$. It is known from the functional analysis we can write the solution (3) concerning functional $I_{\mathbf{p}}[\vec{w}]$ with accuracy up to the first order on small value $w$ in the following form:

$$I_{\mathbf{p}}(\mathbf{x},t)[\vec{w}] =$$
$$exp\left[\sum_i \int d^3 p'\,\beta_i(\mathbf{p},\mathbf{p}')\frac{\delta}{\delta\varepsilon_i(\mathbf{p},\mathbf{p}')}\Bigg|_{\varepsilon_i(\mathbf{p},\mathbf{p}')=0}\right] I_{\mathbf{p}}(\mathbf{x},t)[0] \qquad (4)$$

with the equations for functional derivatives

$$1)\left[\frac{\partial}{\partial t} + \mathbf{v}\nabla + \mu\right]\int d^3 p'\,\beta_i(\mathbf{p},\mathbf{p}')\frac{\delta I_{\mathbf{p}}}{\delta\varepsilon_i(\mathbf{p},\mathbf{p}')}\Bigg|_{\vec{\varepsilon}=\vec{0}}$$
$$= N_i \int d^3 p'\,\beta_i(\mathbf{p},\mathbf{p}')I_{\mathbf{p}'}[0],$$
$$2)\left[\frac{\partial}{\partial t} + \mathbf{v}\nabla + \mu\right]\frac{\delta^2 I_{\mathbf{p}}}{\delta\varepsilon_i(\mathbf{p},\mathbf{p}')\delta\varepsilon_j(\mathbf{p},\mathbf{p}'')}\Bigg|_{\vec{\varepsilon}=\vec{0}} \qquad (5)$$
$$= \left[N_i\frac{\delta I_{\mathbf{p}'}}{\delta\varepsilon_j(\mathbf{p},\mathbf{p}'')} + N_j\frac{\delta I_{\mathbf{p}''}}{\delta\varepsilon_i(\mathbf{p},\mathbf{p}')}\right]\Bigg|_{\vec{\varepsilon}=\vec{0}},$$
$$\cdots$$

which is got from (3). The form of (3) without the scattering term and every equation from (5) are identical. Therefore when we know how to solve (3) without the scattering term so we can solve the chain of the equations from (5).

When we take into account dependence on time only for $F$ (as for example in [5]), kinetic energy of particle depends only on absolute value of impulse that is sufficient the general case so

$$\left[\frac{1}{v(\mathbf{p})}\frac{\partial}{\partial t}+\boldsymbol{\omega}\nabla+\frac{\mu(\mathbf{x};\mathbf{p})}{v(\mathbf{p})}\right]I_{\mathbf{p}}(\mathbf{x},t)=$$

$$\frac{F(\mathbf{x},t;\mathbf{p})}{v(\mathbf{p})}+\int\frac{W(\mathbf{x},t;\mathbf{p},\mathbf{p}')}{v(\mathbf{p})}I_{\mathbf{p}'}(\mathbf{x},t)\,d^3p', \qquad (6)$$

$$v(\mathbf{p})=|\mathbf{v}|=\left|\frac{\partial E}{\partial\mathbf{p}}\right|,\qquad \mathbf{p}/|\mathbf{p}|=\boldsymbol{\omega}.$$

At using boundary conditions $I_{\boldsymbol{\omega}}(\mathbf{x},t)=0$ at $\mathbf{x}\in\partial\Omega$ and $\mathbf{v}\,\boldsymbol{\omega}<0$ ($\mathbf{v}$ is the exterior normal to surface $\partial\Omega$), neglecting scattering the integrating (6) gives:

$$I_{\mathbf{p}}(\mathbf{x},t)[0]=\int\limits_{\mathbf{x}\boldsymbol{\omega}-\infty}^{\mathbf{x}\boldsymbol{\omega}}e^{-\int\limits_{\mathbf{y}}^{\mathbf{x}\boldsymbol{\omega}}\mu_v(s;\mathbf{p})ds}F_v(\mathbf{y},t-t';\mathbf{p})\,dl(\mathbf{y}),$$

$$t'=\frac{\mathbf{x}\boldsymbol{\omega}-\mathbf{y}\boldsymbol{\omega}}{v(\mathbf{p})}, \qquad (7)$$

where $\dfrac{A}{v(\mathbf{p})}=A_v$, $]\mathbf{x}\boldsymbol{\omega}-\infty,\mathbf{x}\boldsymbol{\omega}]$ means the ray at point $\mathbf{x}$ and direct $-\boldsymbol{\omega}$, the integration on $ds,dl$ is yielded along the direction $\boldsymbol{\omega}$. With account of the first order over scattering term we have

$$I_{\mathbf{p}}(\mathbf{x},t)[\overrightarrow{w}]\approx\int\limits_{\mathbf{x}\boldsymbol{\omega}-\infty}^{\mathbf{x}\boldsymbol{\omega}}dl(\mathbf{x}')e^{-\int\limits_{\mathbf{x}}^{\mathbf{x}}\mu_v(s;\mathbf{p})ds}\Big[F_v(\mathbf{x}',t-t';\mathbf{p})+$$

$$\int W_v(\mathbf{x}',t-t';\mathbf{p},\mathbf{p}')d^3p'\int\limits_{\mathbf{x}'\boldsymbol{\omega}'-\infty}^{\mathbf{x}'\boldsymbol{\omega}'}e^{-\int\limits_{\mathbf{x}'}^{\mathbf{x}'}\mu_v(s';\mathbf{p}')ds'}F_v(\mathbf{x}'',t-t'-t'';\mathbf{p}')dl'(\mathbf{x}'')\Big],$$

$$t'=\frac{\mathbf{x}\boldsymbol{\omega}-\mathbf{x}'\boldsymbol{\omega}}{v(\mathbf{p})},\quad t''=\frac{\mathbf{x}'\boldsymbol{\omega}'-\mathbf{x}''\boldsymbol{\omega}'}{v(\mathbf{p}')}, \qquad (8)$$

here the integration on $ds,dl$ and $ds',dl'$ is yielded along the direction $\boldsymbol{\omega}$ and $\boldsymbol{\omega}'$ accordingly.

In general case for the real mediums which we are going to investigate in this paper the functions $N_i$ is stochastic values because of randomness is uncontrolled interactions of considering system as well as information losses about microscopic moving at measurements processes and the initial station of the system. How it is for standard situation when we do not take into account the time dependence in detection process. Thus we can consider (6) as a stochastic equation. At least the two component mediums $\mu(\mathbf{x},t;\mathbf{p})$ and $W(\mathbf{x},t;\mathbf{p},\mathbf{p}')$ are stochastic functions depending on $N_i$. Next we will be more interested in the case when at any parameters the density of radiation sources is either less then the density other absorbing and dissipating medium components, or radiation sources are outside medium bounds. It means that we may neglect by contribution from these sources to scattering and absorption, this means independence on $N_i$ of random components of $F$. Taking to account these views after the averaging (8) over the stochastic functions $N_i$

$$\langle I_{\mathbf{p}}(\mathbf{x},t)[\overrightarrow{w}]\rangle\approx\int\limits_{\mathbf{x}\boldsymbol{\omega}-\infty}^{\mathbf{x}\boldsymbol{\omega}}dl\left\langle e^{-\int\limits_{\mathbf{x}}^{\mathbf{x}}\mu_v(s;\mathbf{p})ds}\right\rangle[F_v(\mathbf{x}',t-t';\mathbf{p})+$$

$$\int d^3p'\int\limits_{\mathbf{x}'\boldsymbol{\omega}'-\infty}^{\mathbf{x}'\boldsymbol{\omega}'}\left\langle W_v(\mathbf{x}',t-t';\mathbf{p},\mathbf{p}')e^{-\int\limits_{\mathbf{x}'}^{\mathbf{x}'}\mu_v(s';\mathbf{p}')ds'}\right\rangle F_v(\mathbf{x}'',t-t'-t'';\mathbf{p}')dl'\Big],$$

$$(9)$$

we can see that essential changing has been taken place in this approach only for term including the attenuation coefficient because in general case an average of exponent function of a stochastic value is not exponent function of the average. Hereinafter we will study what this will give for the reconstruction of $F$ in practice.

Let us continue our investigation of transport model for photons with the energies (below 1 MeV) of interest in nuclear imaging. Usually in such measurements the sources of isotropic radiation are used and we wish to estimate theirs intensity in general together with isotropic attenuation coefficient whereas for transmission statement we reconstruct only isotropic attenuation coefficient. On the base of general transport equation (3) the dynamic transport equation can be written in the following form:

$$\left[\frac{\partial}{\partial t}+\boldsymbol{\omega}\nabla+\mu(\mathbf{x},t;E)\right]I_{\boldsymbol{\omega}}(\mathbf{x},t;E)=F(\mathbf{x},t;E)+$$

$$\sum_{i,j}N_i(\mathbf{x},t)\int\beta_{i,j}(E,\boldsymbol{\omega},\boldsymbol{\omega}')I_{\boldsymbol{\omega}'}\big(\mathbf{x},t;E'_j(E,\boldsymbol{\omega},\boldsymbol{\omega}')\big)\,d\boldsymbol{\omega}', \qquad (10)$$

where it means the propagation velocity of photons is equal to 1. Here $F(\mathbf{x},t;E)$ is the unknown distribution of photons radiated from the point $\mathbf{x}$ and at time $t$, $\mu(\mathbf{x},t;E)$ is the isotropic attenuation coefficient which decomposes for $i^{th}$ component of the medium as

$\mu_i(\mathbf{x},t;E)=N_i(\mathbf{x},t)\sigma_i(E)+N_i(\mathbf{x},t)\sum_j\int\beta_{i,j}(E,\boldsymbol{\omega}',\boldsymbol{\omega})\,d\boldsymbol{\omega}'$ in general case with the unknown density $N_i(\mathbf{x},t)$ of $i^{th}$ component and known terms

$\sigma_i(E)+\sum_j\int\beta_{i,j}(E,\boldsymbol{\omega}',\boldsymbol{\omega})\,d\boldsymbol{\omega}'$ representing all physical

processes which end flights of photons at energy $E$ with initial energy $E_j' \geq E$. For the energy relevant to nuclear imaging these comprise photo effect, Rayleigh scattering and Compton scattering [13]. We may model scattering kernel for $i^{th}$ component and $j^{th}$ scattering mechanism as $W_{i,j}(\mathbf{x},t;E,\boldsymbol{\omega},\boldsymbol{\omega}') = N_i(\mathbf{x},t)\,\beta_{i,j}(E,\boldsymbol{\omega},\boldsymbol{\omega}')$. For example Compton scattering is described the known Klein-Nishina scattering cross section with $E'(E,\boldsymbol{\omega},\boldsymbol{\omega}') = E / \left[1 - \dfrac{E}{m}\boldsymbol{\omega}\boldsymbol{\omega}'\right]$, where $m$ is the electron mass, .

Notice that the energy of photons used in nuclear imagine is often monochromatic but only Rayleigh scattering does not cause energy losses of photons. Assuming that our detector has enough perfect energy resolution and we are capable of recognize photons which have scattered. Rayleigh scattering can be neglected with respect to other scattering mechanism [13]. For monochromatic sources $I_{\boldsymbol{\omega}}(\mathbf{x},t;E) \gg I_{\boldsymbol{\omega}}(\mathbf{x},t;E_j')$ for $E_j' \neq E$. Therefore the scattering term we can consider as perturbation with respect to the term with the attenuation coefficient in (10) since $I_{\boldsymbol{\omega}}(\mathbf{x},t;E)\int \beta_{i,j}(E,\boldsymbol{\omega}',\boldsymbol{\omega})\,d\boldsymbol{\omega}' \gg \int \beta_{i,j}(E,\boldsymbol{\omega},\boldsymbol{\omega}')I_{\boldsymbol{\omega}}(\mathbf{x},t;E_j'(E,\boldsymbol{\omega},\boldsymbol{\omega}'))\,d\boldsymbol{\omega}'$ and (4) are applied. The solution of the simplified dynamic transport equation

$$\left[\frac{\partial}{\partial t} + \boldsymbol{\omega}\nabla + \mu(\mathbf{x},t;E)\right]I_{\boldsymbol{\omega}}(\mathbf{x},t;E) = F(\mathbf{x},t;E) \quad (11)$$

will be initial approach for solution of (10). Importance of the account of the scattering term is greater or lesser depending on monochromaticity degree of sources.

The detector resolution is finite that implies data is collected over energy window $E_0 \pm \Delta E$. Therefore we observe the value

$$I_{\boldsymbol{\omega}}(\mathbf{x},t) = \int_{E_0 - \Delta E}^{E_0 + \Delta E} I_{\boldsymbol{\omega}}(\mathbf{x},t;E)\,dE.$$

In (10) we can take into account this integrating over the energy for window $E_0 \pm \Delta E$ accounting

$$\mu(\mathbf{x},t)\,I_{\boldsymbol{\omega}}(\mathbf{x},t) = \mu(\mathbf{x},t;E*)\int_{E_0 - \Delta E}^{E_0 + \Delta E} I_{\boldsymbol{\omega}}(\mathbf{x},t;E)\,dE,$$

where $E* \in \left[E_0 - \Delta E, E_0 + \Delta E\right]$. Let's note that $\mu(\mathbf{x},t)$ is stochastic function and the more of the independent random factors define the attenuation coefficient the more the probability distribution of stochastic component of function $\mu(\mathbf{x},t)$ is nearer to Gaussian distribution. Finally we can write the following stochastic dynamic transport equation

$$\left[\frac{\partial}{\partial t} + \boldsymbol{\omega}\nabla + \mu(\mathbf{x},t)\right]I_{\boldsymbol{\omega}}(\mathbf{x},t) = F(\mathbf{x},t), \quad (12)$$

where $\mu(\mathbf{x},t)$ and $F(\mathbf{x},t)$ are independent random functions. When we interest only average values of the distribution function from time $t_0$ to $t$ (i.e. $I_{\boldsymbol{\omega}}(\mathbf{x},t_0) \gg I_{\boldsymbol{\omega}}(\mathbf{x},t) - I_{\boldsymbol{\omega}}(\mathbf{x},t_0)$ )

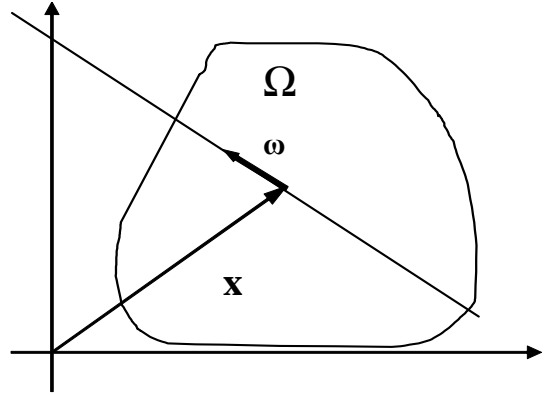

Fig.1. Scheme of measurements and some denotes for CT. $\boldsymbol{\omega}$ is direction in which emitted at $\mathbf{x}$ photons are detected.

after integrating (12) over $t$ we obtain transport equation widespread using in nuclear imaging

$$\left[\boldsymbol{\omega}\nabla + \mu(\mathbf{x})\right]\bar{I}_{\boldsymbol{\omega}}(\mathbf{x}) = \bar{F}(\mathbf{x}),$$

$$\bar{I}_{\boldsymbol{\omega}}(\mathbf{x}) = \frac{1}{\Delta t}\int_{t_0}^{t} I_{\boldsymbol{\omega}}(\mathbf{x},t)\,dt, \; \bar{F}(\mathbf{x}) = \frac{1}{\Delta t}\int_{t_0}^{t} F(\mathbf{x},t)\,dt, \quad (13)$$

$$\mu(\mathbf{x}) \equiv \mu(\mathbf{x},t*), \; t* \in [t,t_0].$$

The omitted time dependence adds an uncertainty to the attenuation coefficient at such measurements.

## III. SPECT, PET AND X-RAY CT

Let us continue our investigation some stochastic aspects of transport model for problems of ECT and X–ray CT. In fact the problem of ECT is the definition of the average density $\rho$ of radiation sources (which does not coincide with $F$ generally speaking) during the projections measurement of distribution function of radiated particles. The random component in $F$ is determined by fluctuations of the number of emitted particles and fluctuations of the density of radiation sources which happen for time less than time of measurement of the given projection. For emission tomography it is clear to suppose that $\rho$ is equal to the averaged $F$ within proportionality factor. The problem of X-ray CT is the definition of average value of random variable $\mu$.

Let's study the problems of ECT and X-ray CT with account of errors in definition $\mu$, statistical heterogeneity of a medium and fluctuations of number of photons which are emitted during the projection measurement. In the static case when at time of projection measurement the average values of random functions are changing small we can use (13). This equation is defined in area $\Omega \subseteq R^n$ for any $\boldsymbol{\omega} \in S^{n-1}$ (see Fig. 1). If we consider a case of absence of radiation sources outside of $\Omega$ :

$$I_\omega\left(\mathbf{x}\right) = 0, \ \mathbf{x} \in \partial\Omega, \ \mathbf{v} \ \boldsymbol{\omega} < 0,$$
$$I_\omega\left(\mathbf{x}\right) = g\left(\boldsymbol{\omega},\mathbf{x}\right), \ \mathbf{x} \in \partial\Omega, \ \mathbf{v} \ \boldsymbol{\omega} \ge 0,$$

where $\mathbf{v}$ is the exterior normal to surface $\partial\Omega$. Without losing generality we can consider the transmission case as distribution sources which locate at points of boundary.

When the function $\mu(\mathbf{x})$ is known the solution of (13) looks like the following in neglecting scattering:

$$I_\omega\left(\mathbf{x}\right) = \int_{\mathbf{x}\omega-\infty}^{\mathbf{x}\omega} \exp\left[-\int_y^x \mu(s)ds\right] F\left(\mathbf{y}\right)dl(\mathbf{y}) \qquad (14)$$

where the integration on $ds, dl$ is yielded along the direction $\boldsymbol{\omega}$.

Thus in case of the statement of ECT problem the random function $g\left(\boldsymbol{\omega},\mathbf{x}\right)$ describing number of photons, which can be detected in a point $\mathbf{x}$ and with velocity in the direction $\boldsymbol{\omega}$, expresses through random functions $F$ and $\mu$

$$g\left(\boldsymbol{\omega},\mathbf{x}\right) = \int_{\mathbf{x}\omega-\infty}^{\mathbf{x}\omega} \exp\left[-\int_y^x \mu ds\right] F\left(\mathbf{y}\right)dl. \qquad (15)$$

When the source emits particles mutually in opposite directions gamma-quanta is detected on coincidence (PET). Then exponent function can be taken out from integral sign

$$g\left(\boldsymbol{\omega},\mathbf{x}\right) = \exp\left[-\int_{\mathbf{x}\omega-\infty}^{\mathbf{x}\omega} \mu ds\right] \int_{\mathbf{x}\omega-\infty}^{\mathbf{x}\omega} F\left(\mathbf{y}\right)dl$$

but for X-ray CT

$$g\left(\boldsymbol{\omega},\mathbf{x}\right) = I_0 \exp\left[-\int_{\mathbf{x}\omega-\infty}^{\mathbf{x}\omega} \mu ds\right]. \qquad (16)$$

Further for using the projection data obtained in practice we have to average the formulas for projections over fluctuations $F(\mathbf{x})$ ( $F\left(x\right) = \overline{F}\left(x\right) + \tilde{F}\left(x\right)$ ) bounded with quantum nature of the radiation process and density fluctuations of radiation sources during measurement, and over random field $\tilde{\mu}(\mathbf{x})$ ( $\mu(x) = \overline{\mu}(x) + \tilde{\mu}(x), \overline{\mu}(x) = \langle\mu(x)\rangle$ ). Taking into account that the average over fluctuations of the attenuation coefficient does not affect $F$ (how we could see above) we shall obtain

$$\boldsymbol{\omega}\nabla\langle I_\omega\rangle + \langle\mu \ I_\omega\rangle = f,$$
$$f = \overline{F}. \qquad (17)$$

Taking into account the averaging over all random parameters we obtain from (14)

$$\langle I_\omega\left(\mathbf{x}\right)\rangle = \int_{\mathbf{x}\omega-\infty}^{\mathbf{x}\omega} \left\langle \exp\left[-\int_y^x \mu(s)ds\right]\right\rangle f\left(\mathbf{y}\right)dl \qquad (18)$$

for ECT and

$$\langle I_\omega\left(\mathbf{x}\right)\rangle = I_0 \left\langle \exp\left[-\int_{\mathbf{x}\omega-\infty}^{\mathbf{x}\omega} \mu\left(s\right)ds\right]\right\rangle \qquad (19)$$

for X-ray CT.

So it is obtained that practically we can reconstruct average density of radiation sources ( $\rho(\mathbf{x}) \approx const \ f\left(\mathbf{x}\right)$ ) in CT after certain statistical processing of detection results which are realization of random function $g\left(\boldsymbol{\omega},\mathbf{x}\right)$. And we have to take into account the distortion of Radon transformation because of the averaging over fluctuations $\tilde{\mu}(\mathbf{x})$ of the factor $\exp[-\int_{\mathbf{x}\omega-\infty}^{\mathbf{x}\omega} \mu ds]$. In the case of PET the factor before $f(\mathbf{x})$ is taken out of integral. But for SPECT and X-ray CT we have something new - in general case (18) and (19) are not reduced to Radon transform. Further we shall study those integral transformations which take into account random components of attenuation coefficient in SPECT, X-ray CT and consequence of this for nuclear imaging.

For full statistic description of random function $\mu$ it is enough to know its characteristic functional:

$$\Phi\left[k\right] = \left\langle \exp\left[i\int_{-\infty}^{+\infty} k\left(x\right)\mu\left(x\right)dx\right]\right\rangle,$$
$$\langle\mu\left(x\right)\rangle = \left.\frac{\delta \ \Phi}{i\delta \ k\left(x\right)}\right|_{k=0}, \qquad (20)$$
$$\langle\mu\left(x\right)\mu\left(x'\right)\rangle = -\left.\frac{\delta^2\Phi}{\delta \ k\left(x\right)\delta \ k\left(x'\right)}\right|_{k=0}.$$

Hence we obtain:

$$\left\langle \exp\left[-\int_y^x \mu\left(s\right)ds\right]\right\rangle_{\tilde{\mu}} = \Phi[k]\big|_{k=i\theta(s;\mathbf{x}\omega,\mathbf{y}\omega)}, \quad \theta\left(s;\mathbf{x}\omega,\mathbf{y}\omega\right) \qquad (21)$$
$$= \delta(\mathbf{x}^\perp - \mathbf{y}^\perp)\{\theta(\mathbf{x}\omega - s) - \theta(\mathbf{y}\omega - s)\}, s = \mathbf{y}'\boldsymbol{\omega},$$

here and further $\boldsymbol{a}^\perp$ denotes the projection of vector $\boldsymbol{a}$ onto plane which is perpendicular to $\boldsymbol{\omega}$.

So, it is possible to rewrite (18) in the following form

$$\left\langle I_{\omega}\left(\mathbf{x}\right)\right\rangle_{\tilde{\mu}} = \int\limits_{\mathbf{x}\omega-\infty}^{\mathbf{x}\omega} \Phi[i\theta\left(s;\mathbf{x}\omega,\mathbf{y}\omega\right)]f\left(\mathbf{y}\right)dl,$$

$$\left\langle \mu(\mathbf{x})I_{\omega}\left(\mathbf{x}\right)\right\rangle_{\tilde{\mu}} = \int\limits_{\mathbf{x}\omega-\infty}^{\mathbf{x}\omega} \frac{\delta\Phi[k]}{i\delta k}\bigg|_{k=i\theta(s;\mathbf{x}\omega,\mathbf{y}\omega)} f\left(\mathbf{y}\right)dl \qquad (22)$$

Hereinafter under $I_{\omega}$ and $g\left(\omega,\mathbf{x}\right)$ we understand the values averaged over fluctuations $\tilde{F}$. It is necessary to note, that when we can construct characteristic functional for the random function $\mu(\mathbf{x})$, (22) is convenient for practical definition $f(\mathbf{x})$ using observable $\left\langle g\left(\omega,\mathbf{x}\right)\right\rangle_{\tilde{\mu}}$.

At the same time (17) can be written by the following form:

$$\omega\nabla\left\langle I_{\omega}\right\rangle_{\tilde{\mu}} + \bar{\mu}\left\langle I_{\omega}\right\rangle_{\tilde{\mu}} + \left\langle\tilde{\mu}\quad I_{\omega}\right\rangle_{\tilde{\mu}} = f \qquad (23)$$

With account of the well known property

$$\left\langle \tilde{k}(l)\Psi\left[\bar{k}+\tilde{k}\right]\right\rangle_{\tilde{k}} = \Lambda\left[l;\frac{\delta}{i\delta\bar{k}}\right]\left\langle \Psi\left[\bar{k}+\tilde{k}\right]\right\rangle_{\tilde{k}},$$

$$\Lambda[l;q(s)] = \frac{\delta}{i\delta q(l)}\ln\Phi[q(s)] \qquad (24)$$

we obtain the following relation

$$\left\langle \tilde{\mu}\left(\mathbf{x}\right)I_{\omega}\left(\mathbf{x}\right)\right\rangle_{\tilde{\mu}} = \left\langle\Lambda\left[\mathbf{x};\frac{\delta}{i\delta\bar{\mu}\left(\mathbf{x}'\right)}\right]I_{\omega}\left(\mathbf{x}\right)\right\rangle_{\tilde{\mu}}, \qquad (25)$$

which allows to exclude random function $\tilde{\mu}$ from (23),

$$\omega\nabla\left\langle I_{\omega}(\mathbf{x})\right\rangle_{\tilde{\mu}} + \bar{\mu}(\mathbf{x})\left\langle I_{\omega}(\mathbf{x})\right\rangle_{\tilde{\mu}} + \left\langle\Lambda\left[\mathbf{x};\frac{\delta}{i\delta\bar{\mu}(\mathbf{x}')}\right]I_{\omega}(\mathbf{x})\right\rangle_{\tilde{\mu}} = f(\mathbf{x}). \quad (26)$$

In particular for Gaussian stochastic function $\tilde{\mu}$ we get

$$\omega\nabla\left\langle I_{\omega}(\mathbf{x})\right\rangle + \bar{\mu}(\mathbf{x})\left\langle I_{\omega}(\mathbf{x})\right\rangle + \int\limits_{-\infty}^{+\infty}\left\langle\tilde{\mu}(\mathbf{x})\tilde{\mu}(\mathbf{x}')\right\rangle\frac{\delta\left\langle I_{\omega}(\mathbf{x})\right\rangle}{\delta\tilde{\mu}(\mathbf{x}')}d^n x' = f(\mathbf{x}), (27)$$

here and further we omit index $\tilde{\mu}$ in sign of the averaging.

Thus after the averaging of (13) we have got integro-differential equation that means presence coordinate derivatives of order n>1 in the equation for $\left\langle I_{\omega}\right\rangle$ and consequently the dissipation. Besides there is the renormalization of attenuation coefficient and the particles velocity, that corresponds to zero and first derivation accordingly.

## IV. GAUSSIAN NOISE

Let us consider that attenuation coefficient $\mu$ is involved with fluctuations which are called Gaussian noise. The distributions with Gaussian characteristic functional meet in many situations. We shall use the most common form of Gaussian characteristic functional [14] for $\tilde{\mu}(x)$:

$$\Phi[k] = \exp\left[-\frac{1}{2}\iint k\left(x\right)k\left(x'\right)A\left(x,x'\right)dxdx'\right], \qquad (28)$$

where $A(x,x') = \left\langle\tilde{\mu}(x)\tilde{\mu}(x')\right\rangle$ is correlation function. In concrete cases the form of the function $A$ can be determined on the base of the physical situation. So, from (22) we have

$$\left\langle I_{\omega}(\mathbf{x})\right\rangle = \int\limits_{\mathbf{x}\omega-\infty}^{\mathbf{x}\omega} \exp\left[\frac{1}{2}\iint\limits_{\mathbf{y}}^{\mathbf{x}\,\mathbf{x}} A_{\omega}\left(s,s'\right)dsds' - \int\limits_{\mathbf{y}}^{\mathbf{x}}\bar{\mu}(s)ds\right]f(\mathbf{y})dl(\mathbf{y}), (29)$$

where

$$A_{\omega}\left(s,s'\right) = \int A\left(\mathbf{z}',\mathbf{z}''\right)\delta(\mathbf{x}^{\perp}-\mathbf{z}'^{\perp})\delta(\mathbf{x}^{\perp}-\mathbf{z}''^{\perp})d^{n-1}\mathbf{z}'^{\perp}d^{n-1}\mathbf{z}''^{\perp},$$

$$\mathbf{x}^{\perp}\omega = \mathbf{z}'^{\perp}\omega = \mathbf{z}''^{\perp}\omega = 0, \quad s = \mathbf{z}'\omega - \mathbf{y}\omega, \quad s' = \mathbf{z}''\omega - \mathbf{y}\omega.$$

Further using (26) and (29) we obtain

$$\omega\nabla\left\langle I_{\omega}(\mathbf{x})\right\rangle + \bar{\mu}(\mathbf{x})\left\langle I_{\omega}(\mathbf{x})\right\rangle = f(\mathbf{x})$$
$$+ \int\limits_{-\infty}^{+\infty}\left\langle\tilde{\mu}(\mathbf{x})\tilde{\mu}(\mathbf{x}')\right\rangle\theta(\mathbf{x}\omega-\mathbf{x}'\omega)\delta(\mathbf{x}^{\perp}-\mathbf{x}'^{\perp})\left\langle I_{\omega}(\mathbf{x})\right\rangle d^n x'. \qquad (30)$$

It is well visible that the presence of Gaussian noise component causes the appearance of (after the averaging in right side of (13) ) the term which has the same functional form as scattering term from (3). From (29) it is possible to conclude that we can restore $f(\mathbf{x})$ even at locally great magnitude of Gaussian noise if the correlation function $A_{\omega}\left(s,s'\right)$ is known.

Let's illustrate the influence of random components on the reconstruction on example of the simple asymptotic case called Gaussian Markov process ("colored" noise):

$$\left\langle\tilde{\mu}(\mathbf{x})\right\rangle = 0, \ A(\mathbf{x};\mathbf{x}') = \left\langle\tilde{\mu}(\mathbf{x})\tilde{\mu}(\mathbf{x}')\right\rangle = \sigma^2\exp[-\alpha|\mathbf{x}-\mathbf{x}'|],$$
$$\sigma^2 = h\alpha, \quad \lim_{\alpha\to\infty}\sigma^2\exp[-\alpha|\mathbf{x}-\mathbf{x}'|] = 2h\delta(\mathbf{x}-\mathbf{x}'), \qquad (31)$$

where $\sigma$ is dispersion of $\tilde{\mu}(x)$, $0\le h<\infty$, $\frac{1}{\alpha}$ is radius correlations. At $\alpha\to\infty$, we obtain from (30) the equation with the terms of the order which is no less than $\frac{1}{\alpha^2}$,

$$-\frac{h}{\alpha^2}(\boldsymbol{\omega}\nabla)^2\langle I_\omega(\mathbf{x})\rangle+(1-\frac{h}{\alpha})\boldsymbol{\omega}\nabla\langle I_\omega(\mathbf{x})\rangle+(\bar{\mu}(\mathbf{x})-h)\langle I_\omega(\mathbf{x})\rangle=f(\mathbf{x}). \quad (32)$$

Thus the dissipation (first term (32)) has been appeared. We can see that the renormalization of the attenuation coefficient to $\bar{\mu}-h$ and the velocity to $1-h/\alpha$ takes place in (32). If $\alpha>1/\Delta$, where $\Delta$ is necessary resolution (i.e. radius of correlation tend to zero and in the limit the "colored" noise becomes "white" when radius of correlation is much less of characteristic distances in the system) and $h/\alpha\ll1$ we obtain the situation that been reduced to attenuated Radon transformation with renormalized attenuation coefficient $\bar{\mu}\to\mu^*=\bar{\mu}-h$. At other values of $\alpha$ and $h$ we have the case which is not reduced to Radon transformation with account of the attenuation. Further we will investigate this situation in details.

From (29) we obtain

$$\langle g(\boldsymbol{\omega},\mathbf{x})\rangle=\int\limits_{\mathbf{x}\boldsymbol{\omega}=-\infty}^{\mathbf{x}\boldsymbol{\omega}}\exp\left\{h[\tau(\boldsymbol{\omega},\mathbf{x}^\perp)-\mathbf{y}\boldsymbol{\omega}+\frac{1}{\alpha}(e^{-\alpha(\tau(\boldsymbol{\omega},\mathbf{x}^\perp)-\mathbf{y}\boldsymbol{\omega}})-1)]\right\}$$
$$\times e^{-\int\limits_\mathbf{y}^\mathbf{x}\bar{\mu}(s)ds}f(\mathbf{y})\,dl=\int\limits_{\mathbf{x}\boldsymbol{\omega}=-\infty}^{\mathbf{x}\boldsymbol{\omega}}\exp\left\{\frac{h}{\alpha}(e^{-\alpha(\tau(\boldsymbol{\omega},\mathbf{x}^\perp)-\mathbf{y}\boldsymbol{\omega}})-1)]\right\}$$
$$\times e^{-\int\limits_\mathbf{y}^\mathbf{x}\mu^*(s)ds}f(\mathbf{y})\,dl, \quad (33)$$

here $\mathbf{x}^\perp+\tau(\boldsymbol{\omega},\mathbf{x}^\perp)\boldsymbol{\omega}$ is cross point of the ray, which is going out from $\mathbf{x}^\perp$ in the direction $\boldsymbol{\omega}$, with the boundary of the area $\Omega$ at condition $\boldsymbol{\nu}\cdot\boldsymbol{\omega}\geq0$. Let's remark that $\tau(\boldsymbol{\omega},\mathbf{x}^\perp)-\mathbf{y}\boldsymbol{\omega}\geq0$ thus factor $\exp\left\{\frac{h}{\alpha}(e^{-\alpha(\tau(\boldsymbol{\omega},\mathbf{x}^\perp)-\mathbf{y}\boldsymbol{\omega}})-1)\right\}$ (the cause of the difference of (33) from Radon transform with the attenuation) from the core of the integral transformation is less or equal to 1. Here it is necessary to note that having respect to physical sense the inequality $\int\limits_\mathbf{y}^\mathbf{x}\mu^*(s)\,ds\geq0$ has to be true.

In case of X-ray CT

$$\langle g(\boldsymbol{\omega},\mathbf{x})\rangle=I_0\exp\left\{\frac{h}{\alpha}(e^{-\alpha L(\boldsymbol{\omega},\mathbf{x}^\perp)}-1)]\right\}e^{-\int\limits_{\mathbf{x}\boldsymbol{\omega}=-\infty}^{\mathbf{x}\boldsymbol{\omega}}\mu^*(s)ds} \quad (34)$$

where $L(\boldsymbol{\omega},\mathbf{x}^\perp)$ is the size of $\Omega$ at $\mathbf{x}^\perp$ in the direction $\boldsymbol{\omega}$. Using (34) we have to correct the projections then different methods for the reconstruction can be applied as well as it is in the general case for PET.

How we can see from (19) and (29) in case of Gaussian noise for X-ray CT the inverse Radon transform with the correction of projections is the solution of (19) with respect to renormalized average attenuation coefficient $\mu^*$. Whereas for SPECT on this way we have obtained the elementary conditions for the noise parameters limiting the application of Radon transformation with the attenuation without the account of stochastic phenomena: the correlation radius has to be less than necessary resolution, the square of the product of noise dispersion by the correlation radius has to be much less than one and in turn the inequality $\int\limits_\mathbf{y}^\mathbf{x}\mu^*(s)\,ds\geq0$ has to be true for self-consistency of the approach.

## V. POINT SOURCE

The influence of errors upon reconstruction at the definition of the projections is investigated quite [3], [9]. Let's consider the image reconstruction at the presence of "colored" noise in attenuation coefficient on example of the point source with intensity $I(d(\mathbf{y})=I\delta(\mathbf{y}-\mathbf{y}_0))$. We shall estimate the errors bounded with unaccounted noise component in the case of inverse exponential Radon transformation used widely in practice, which in our approach correspond to renormalized averaged coefficient $\mu^*$. Let's stop on the factor $\exp\left\{\frac{h}{\alpha}(e^{-\alpha(\tau(\boldsymbol{\omega},\mathbf{x}^\perp)-\mathbf{y}\boldsymbol{\omega}})-1)\right\}$ from the core of integral transformation (33) which is the reason of difference (33) from exponential Radon transformation. In this case for $\Omega\subset R^2$ we have for averaged projection data from (33)

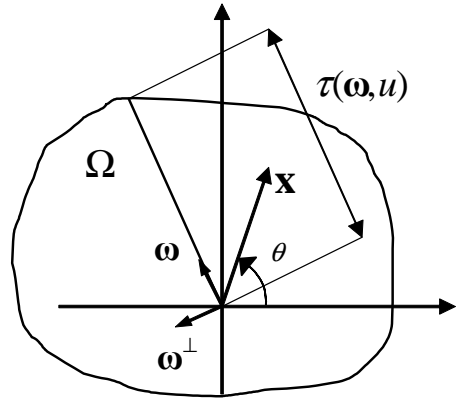

Fig. 2 Coordinate system and parameters use in calculations.

$$\bar{g}(\boldsymbol{\omega},u)=I\delta(u-\boldsymbol{\omega}^\perp\mathbf{y}_0)G(\boldsymbol{\omega})e^{-\mu^*(\tau(\boldsymbol{\omega},\mathbf{y}_0\boldsymbol{\omega}^\perp)-\mathbf{y}_0\boldsymbol{\omega})},$$
$$G(\boldsymbol{\omega})=\exp\left\{\frac{h}{\alpha}(e^{-\alpha(\tau(\boldsymbol{\omega},\mathbf{y}_0\boldsymbol{\omega}^\perp)-\mathbf{y}_0\boldsymbol{\omega})}-1)\right\}, \quad (35)$$

here $u\boldsymbol{\omega}^\perp+\tau(\boldsymbol{\omega},u)\boldsymbol{\omega}$ is cross point with the boundary of the area $\Omega$ of the ray which is going out from $u\boldsymbol{\omega}^{\perp}$ in the direction $\boldsymbol{\omega}$ (see Fig.2).

In this case the inverse transformation for exponential Radon transformation looks like as the following [9]

$$R_{\mu^*}^{-1}(\mathbf{x})\{e^{\mu^*\tau\left(\boldsymbol{\omega},\mathbf{y_0}\boldsymbol{\omega}^{\perp}\right)}\overline{g}(\boldsymbol{\omega},u)\}=\tilde{d}(\mathbf{x}),$$

$$R_{\mu^*}^{-1}(\mathbf{x})\{e^{\mu^*\tau\left(\boldsymbol{\omega},\mathbf{y_0}\boldsymbol{\omega}^{\perp}\right)}\overline{g}(\boldsymbol{\omega},u)\}$$

$$=IN\int_{S^1}e^{-\mu^*\mathbf{x}\boldsymbol{\omega}}d\boldsymbol{\omega}\int_{\mu\le|\eta|<+\infty}e^{i\eta\mathbf{x}\boldsymbol{\omega}^{\perp}}\hat{g}(\boldsymbol{\omega},\eta)|\eta|\,d\eta,$$

$$\hat{g}(\boldsymbol{\omega},\eta)=\frac{1}{\sqrt{2\pi}}\int_{-\infty}^{\infty}e^{-i\eta u}\overline{g}(\boldsymbol{\omega},u)du,\ \ N=\frac{1}{2}(2\pi)^{-1}. \tag{36}$$

It is well known that the different filters of low frequency $\Phi(\eta)$ [9] are used for the algorithms of FBP. Taking into account it we obtain for point source

$$\tilde{d}_b(\mathbf{x})=IN\times$$

$$\int_{S^1}e^{-\mu^*(\mathbf{x}-\mathbf{y_0})\boldsymbol{\omega}}d\boldsymbol{\omega}\int_{\mu\le|\eta|<\infty}e^{i\eta(\mathbf{x}-\mathbf{y_0})\boldsymbol{\omega}^{\perp}}G(\boldsymbol{\omega})\Phi(\eta)d\eta, \tag{37}$$

$$\lim_{b\to\infty}\tilde{d}_b(\mathbf{x})=\tilde{d}(\mathbf{x}),$$

where $0\le\Phi(\eta)\le\dfrac{1}{\sqrt{2\pi}}|\eta|$ and $\Phi(\eta)=0$ for $|\eta|>b$.

Since $0<G(\boldsymbol{\omega})\le1$ this integral satisfies the following inequality

$$|\tilde{d}_b(\mathbf{x})|\le IN|\int_{S^1}e^{-\mu^b(\mathbf{x}-\mathbf{y_0})\boldsymbol{\omega}}d\boldsymbol{\omega}\int_{\mu\le|\eta|\le\infty}e^{i\eta(\mathbf{x}-\mathbf{y_0})\boldsymbol{\omega}^{\perp}}\Phi(\eta)d\eta|$$

$$\le I|\delta^b(\mathbf{x}-\mathbf{y_0})|,\lim_{b\to\infty}\delta^b(R)=\delta(R), \tag{38}$$

here $\delta^b(R)=(2\pi)^{-1}bJ_1(bR)/R$, $|\mathbf{x}-\mathbf{y_0}|=R$. To calculate $\tilde{d}_b(\mathbf{x})$ at $\Phi(\eta)=\dfrac{1}{\sqrt{2\pi}}|\eta|$ and $\Phi(\eta)=0$ for $|\eta|>b$ let's consider (37) in coordinate system (see.Fig.2) where

$$(\mathbf{x}-\mathbf{y_0})\boldsymbol{\omega}=\mathbf{x}\boldsymbol{\omega}=R\cos(\theta-\varphi),\quad(\mathbf{x}-\mathbf{y_0})\boldsymbol{\omega}^{\perp}=\mathbf{x}\boldsymbol{\omega}^{\perp}=R\sin(\theta-\varphi)$$

$$\boldsymbol{\omega}=(\cos\varphi,\sin\varphi),\quad\mathbf{x}-\mathbf{y_0}=(R\cos\theta,R\sin\theta).$$

First we shall integrate over $\eta$ using identity

$$e^{-\mu R\sin(\theta-\varphi)+i\eta R\cos(\theta-\varphi)}$$

$$=\sum_n i^n r^n J_n(R\tilde{\eta})e^{in(\theta-\varphi)}, \tag{39}$$

where $r=\left(\dfrac{\eta+\mu}{\eta-\mu}\right)^{1/2},\eta=\sqrt{\tilde{\eta}^2+\mu^2}$. Further, in (37) the integrals appear the following types:

$$K_n(R,\mu;b)=\int_\mu^b\frac{(\mu+\eta)^n+(\mu-\eta)^n}{\tilde{\eta}^n}J_n(R\tilde{\eta})\eta\,d\eta, \tag{40}$$

it can seem surprising which are explicitly integrable with Bessel functions. At $b>>1$ we obtain

$$K_{2m-1}(R,\mu;b)=K_m^{(1)}(R,\mu)+O(1/\sqrt{b}),$$

$$K_{2m}(R,\mu;b)=(-1)^m4\pi\delta^b(R)+K_m^{(2)}(R,\mu)+O(1/\sqrt{b}), \tag{41}$$

where m=1,2… and

$$K_m^{(1)}(R,\mu)=2(2m-1)\frac{\mu}{R}+\mu^3R\sum_{k=1}^{m-1}c_{m,k}^{(1)}(\mu R)^{2(k-1)},$$

$$K_m^{(2)}(R,\mu)=4m\frac{1}{R^2}+\mu^2\sum_{k=0}^{m-1}c_{m,k}^{(2)}(\mu R)^{2k},$$

$$c_{m,k}^{(1)}=\frac{2^{1-2k}}{(m+k-1)!}\sum_{j=m-k-1}^{m-1}\frac{j!(2m-1)!}{(j+k+1-m)!(2j)!(2m-1-2j)!},$$

$$c_{m,k}^{(2)}=\frac{2^{-2k}}{(m+k)!}\sum_{j=m-k-1}^{m}\frac{j!(2m)!}{(j+k+1-m)!(2j)!(2m-2j)!}.$$

With account of (41) reconstructing function of point source at $\mathbf{y_0}$ is

$$\tilde{d}_b(\mathbf{x})=I[G_0+2\sum_{m=1}\mathrm{Re}(e^{i2m\theta}G_{2m})]\delta^b(R)+$$

$$\sum_{m=1}(-1)^{m+1}K_m^{(1)}(R,\mu)\,\mathrm{Im}(e^{i(2m-1)\theta}G_{2m-1})+$$

$$\sum_{m=1}(-1)^mK_m^{(2)}(R,\mu)\,\mathrm{Re}(e^{i2m\theta}G_{2m})+O(1/\sqrt{b}), \tag{42}$$

$$G_n=\frac{1}{2\pi}\int_0^{2\pi}G(\boldsymbol{\omega})e^{-in\varphi}d\varphi.$$

As $\lim_{n\to\infty}K_n(R,\mu;b)=0$ at least for $0<Rb<n$ and the Fourier coefficients decrease fast with $n$ (at least as $o(1/n)$) so the part of (42) which does not dependence on $b$ is the limited function. Therefore this part of (42) is equal to zero taking into account (38).

Now with account of

$$\sum_{m=1}^{\infty} \mathrm{Re}(e^{j2m\theta}G_{2m}) = \sum_{m=1}^{\infty} \frac{1}{2\pi} \int_0^{2\pi} G(\omega)\cos[2m(\theta-\varphi)]d\varphi = \frac{1}{4}[G(\theta)+G(\theta+\pi)] - \frac{1}{2}G_0$$

we obtain for $\tilde{d}_b$ at $b \gg 1$

$$\tilde{d}_b(\mathbf{x}) = I\frac{1}{2}[G(\theta)+G(\theta+\pi)]\delta^b(R) + O(1/\sqrt{b}). \quad (43)$$

So, we can see the decreasing and depending on angle $\theta$ of radiation intensity of the point source in comparison with the initial which is equal to $I$. In addition the cases are possible when at definite angles the intensity of reconstructed source is strongly reduced that evidently causes fluctuations of brightness on reconstructing image.

It is well known [9] if the noise is not taken into account at finite $b$ the reconstructed point source has width of order $2\pi/b$ ($b \gg 1$) this is determined by the first term in (43). But with account of the noise the additional widening of reconstructed function appears because of second term in (43), which at $R \approx \pi/b$ is equal to $I\,M(\theta)b^2$ ($M(\theta)<1$) on an order of magnitude. So the contribution to $\tilde{d}$ of the first term (in decreasing order of Fourier coefficients) is equal to $-I\,0.12b^2\,\mathrm{Re}(e^{j2\theta}G_2)$ (see (41) and (42)). That is comparable value with the maximum of the first term of (43) at $R=0$, which is equal to $I\frac{1}{2}[G(\theta)+G(\theta+\pi)]b^2/4\pi$.

Let's pay attention at the fact that because of the noise at using of the method of inverse projection we can reliably restore minimum details by size $D(\theta)>2\pi/b$ depending on the angle. There are the best of the resolution for central area and degradation to edge of investigated area for this kind of a noise. In addition the maximal resolution $D \approx 2\pi/b$ is possible only when the investigated area is circle with radiating source in centre. While at realization of a Nyquist condition (at the specified value of discretization step $h \le \pi/b$) in principle it is possible to reconstruct details of size $2\pi/b$ [9] independently of the place of radiating source. It is clear that at the presence of the set of point sources the interference arises because of the dependence on angle causing the structure distortion of image details and the resolution degradation. In some cases the appearance of artifacts is possible.

## VI. Conclusion

We investigated the transport equation when its solution can be expressed by expansion into a functional Taylor series in terms of the scattering functions. The functional derivatives are determined from the chain of coupled equations which have the similar type. Using dynamic attenuated Radon transform the every functional derivative is determined from the inhomogeneous term which is derived from the previous equation. In detail we considered first-order correction on the scattering to the standard Radon transforms and the changing of the form attenuated Radon transform after the averaging on the random values. It has been shown that neglecting the scattering we can consider attenuation coefficient and the distribution of emitted particles as the independent stochastic functions.

For the case photons with the energies (below 1 MeV) of interest in nuclear imaging the conditions has been analyzed when we can neglect scattering term in the stochastic integral equation, that is satisfied in most conditions for statements of the CT problems in consequence of the monochrome source of photons. It is interesting that averaging of these equations over time smoothes out the fluctuations of the flow density of detected particles and the density of radiation sources, but adds uncertainty in attenuation coefficient. So here is shown, that at the define conditions using the reconstruction algorithms it is important to take into account primarily stochastic nature of attenuation coefficient to obtain necessary quality of image. The equation (9) provides a good perspective for the image reconstruction with account of scattering. In addition it is necessary to investigate $I_\mathbf{p}(\mathbf{x},t)$ with account of the contributions of next orders of functional derivatives with respect to the scattering function from (3) and to describe the conditions of staging of problems of image reconstruction, when it is important in applications.

In this way studying nuclear imaging we have obtained that in SPECT, PET or X-ray CT one deals with some effective attenuation coefficient $\mu^*$ which looks like $\mu^* = \bar{\mu} - h$ in the elementary case of "white" noise. Besides additionally for "color" noise there are effective velocity $(1-h/\alpha)$ of photons and dissipation in (32). We have shown for ECT that it is practically useful to divide attenuation coefficient on weakly changing function in comparison with a constant in a known region (or in general case the function having beforehand guessed form) and the small addend considered as the noise which parameters can be experimentally or theoretically determined. Perhaps it is quite enough to estimate both dispersion and correlation radius for practical applications.

Within the framework of the simplest transport model (see (13)) for PET the account of stochastic aspects can only result in the correction of projections as well as for X-ray CT in the case of the presence of Gaussian noise in attenuation coefficient. Whereas for SPECT with presence of Gaussian noise in attenuation coefficient the correctness of the inverse Radon transform is defined the following conditions: the correlation radius is less than necessary resolution and the square of the product of noise dispersion by correlation radius is much less than one. In other cases the deviation of the actual mechanism of the generation of emission data from Radon transformation with the attenuation causes the sensible difference of the reconstructed from real density of radiation sources in SPECT. If the approximation of Gaussian noise is not true these differences can be essential for X-ray CT too. Besides the unaccounted noise component in attenuation coefficient causes the distortion of image texture and degrades the resolution depending on the boundary form of investigation area and may be the source of artifacts in some cases.

For the further solution of CT problems, it is perspective to study methods of corrective actions on the basis of modification of the integral transformation for SPECT (15) and for X-ray CT (16) to improving the reconstruction methods being based on inverse integral transformation. In other side we can make corrections evidently knowing the expressions for the distortion of reconstructed function for point source such as (43) when the approximation of random components of attenuation coefficient by Gaussian Markov random field is correct. By the same way of the studying of the modification of other methods of image reconstruction is interesting with account of above mentioned in this paper.

It is interesting that there is explicit expression for the integrals in (40). I do not find any consideration of this integrals family elsewhere.

### Acknowledgment

The author would like to thank V.V. Yanovskiy for his contribution to the problem definition and stimulating discussions.